\documentclass[12pt]{article}
\usepackage{amsmath,amsfonts,amssymb,latexsym,graphicx}
\DeclareGraphicsExtensions{.pdf,.jpg}

\numberwithin{equation}{section}
\def\be{\begin{equation}}
\def\ee{\end{equation}}
\def\bq{\begin{eqnarray}}
\def\eq{\end{eqnarray}}
\def\beq{\begin{eqnarray}}
\def\eeq{\end{eqnarray}}

\begin{document}
%%%%%%%%%%%%%%%%%%%%%%%%%%%%%%%%%%%%%%%%%%%%%%%%%%
%%%%%%%%%%%%%%%%%%%%%%%%%%%%%%%%%%%%%%%%%%%%%%%%%%
%%%%%%%%%%%%%%%%%%%%%%%%%%%%%%%%%%%%%%%%%%%%%%%%%%

\title{\Large{\textsc{Curved branes with regular support}}}
\author{{\large\textsc{Ignatios Antoniadis$^{1,2}$\thanks{\texttt{antoniad@lpthe.jussieu.fr}}, Spiros Cotsakis\footnote{On leave from the University of the Aegean, 83200 Samos, Greece.}\,\,$^{3}$\thanks{\texttt{skot@aegean.gr}}}}, %\\
{\large\textsc{Ifigeneia Klaoudatou$^{3}$\thanks{\texttt{iklaoud@aegean.gr}},}} \\[10pt]
$^1$LPTHE, UMR CNRS 7589, Sorbonne Universit\'es, UPMC Paris 6,\\
4 place Jussieu, T13-14, 75005 Paris, France\\
$^2$ Albert Einstein Center for Fundamental Physics, ITP,\\
University of Bern, %\\
Sidlerstrasse 5 CH-3012 Bern, Switzerland\\
$^{3}$Department of Mathematics, %\\
American University of the Middle East\\
P. O. Box 220 Dasman, 15453, Kuwait }

\maketitle
\begin{abstract}
\noindent
We study spacetime singularities in a general five-dimensional braneworld with curved branes
satisfying four-dimensional maximal symmetry. The bulk is supported by an
analog of perfect fluid with the time replaced by the extra coordinate. We show that contrary
to the existence of finite distance singularities from the brane location in any solution with
flat (Minkowski) branes, in the case of curved branes there are singularity-free solutions for a range of equations of state compatible with the null energy condition.

\end{abstract}
\section{Introduction}
In previous work we studied the singularity structure of a braneworld model consisting of a flat
$3$-brane embedded in a five-dimensional bulk space filled with an analogue of a perfect fluid
(the fifth coordinate $Y$ playing the role of time). The perfect fluid satisfied a
linear equation of state with a constant parameter $\gamma$, $P=\gamma\rho$, where $P$ is the `pressure'
and $\rho$ is the `density'. In \cite{ack3} we showed that for a flat brane there exist
singularities that appear within finite distance $Y_{0}$ from the position of the brane
supposedly located at the origin, for all values of $\gamma$. A way to avoid such singularities is to exploit the natural $Z_2$ symmetry
introduced by the existence of the brane by cutting the bulk space and considering a slice of it which
%and matching the part of the bulk space that
is free from finite-distance singularities. Although this matching mechanism is possible for
all values of $\gamma$, the requirement for localised gravity on the brane restricts
$\gamma$ in the interval $(-2,-1)$. On the other hand, further requirements for physical conditions,
such as energy conditions, restrict $\gamma$ in values greater than $-1$. Therefore within the
framework of our flat brane model it is not possible to satisfy at the same time the positive energy conditions and the condition for localised gravity on the brane.

A question that naturally arises is whether any of the above conclusions about the existence of singularities are
sensitive to the geometry of the brane so that singularities are absent when we consider a curved brane.
%Supporting indications for the avoidance of finite-distance singularities for a curved brane were first given in \cite{gubser}.
%In \cite{ack1} and \cite{ack2} we studied this possibility by performing an asymptotic
%analysis method and found various asymptotic behaviours that implied that finite-distance singularities
%where shifted to infinite distance for the case of a curved brane.
It was proposed in~\cite{ack1}, \cite{ack2} that the singularity present in the flat
brane model moves to infinite distance when the brane becomes curved, such as de Sitter (dS)
or anti-de Sitter (AdS) in the maximally symmetric case, which was in accordance with previous claims
made in \cite{gubser}.
%putting aside of course the fine-tuning problem of the four-dimensional (4d) cosmological constant
%and as it was shown such branes offer the possibility of putting aside of course the fine-tuning problem of the four-dimensional (4d) cosmological constant.
%Indeed, as it followed by a detailed asymptotic analysis performed in~\cite{ack1}, \cite{ack2} that the singularity present in the flat brane model
%moves to infinite distance when the brane becomes curved. The simplest examples of curved
%branes include de Sitter (dS) or anti-de Sitter (AdS) in the maximally symmetric case,
%and as it was shown such branes offer the possibility of putting aside of course the fine-tuning problem of the four-dimensional (4d) cosmological constant.
%Supporting indications for the avoidance of finite-distance singularities for a curved brane were first given in \cite{gubser}.
%In \cite{ack1} and \cite{ack2} we studied this possibility by performing an asymptotic
%analysis method and found various asymptotic behaviours that implied that finite-distance singularities
%where shifted to infinite distance for the case of a curved brane.
%Of course, the cost of replacing the flat brane with a curved one the reemerging of the cosmological constant problem.

In this paper, we show that this is indeed possible. In particular, we show that
for curved branes there exist ranges of $\gamma$ for which finite-distance
singularities are avoided; these are: $\gamma>-1/2$ (for positively curved brane) and
$-1<\gamma<-1/2$ (for negatively curved brane). For each type of brane geometry and values of
$\gamma$ outside these regions, we find that finite-distance singularities continue to exist.
A way to be removed is by using the cutting and matching procedure mentioned above to construct a slice of non-singular bulk space, when that is possible.
Moreover, imposing the null energy condition (guaranteeing the absence of ghosts in the bulk)
excludes de Sitter (dS) branes and one is left only with the second region of $\gamma$ for
Anti-de Sitter (AdS) branes.
However, we further show that this region in AdS branes is incompatible with having also localised gravity on the brane.
The situation is not improved when allowed non-singular solutions obtained by the cutting and matching procedure.
%Imposing further conditions such as the localisation of gravity and the energy conditions, we
%find that for the latter matched solutions it \emph{does} become possible to satisfy both
%conditions for $-2<\gamma<-1$.

%It is amusing to study the interplay between the distance from the brane location of the singularity and the value of the curvature of the brane. {\bf can we study this?}

The plan of this paper is the following: In Sections $2$ and $3$, we present the model and give the exact
solutions respecting 4d maximal symmetry,
as well as the complete list of all asymptotic behaviours for all ranges
of the parameters in our model. In Section $4$, we analyse in detail the
two non-singular solutions found in Section $3$. In Section $5$, we derive the
null energy condition and investigate its consequences. In Section $6$ we construct
non-singular orbifold-like solutions, called matching in the following, obtained by the cutting
and matching procedure for those cases that allow it, and study the null energy condition.
In Section $7$, we examine whether the previously derived solutions also satisfy the condition for
localisation of gravity on the brane. Section $8$ contains a summary of our results
and some concluding remarks. Finally, in Appendix A we derive the solutions found for two special
values of the parameter $\gamma$ that cannot be incorporated in the solutions given in
Section $3$.
%%%%%%%%%%%%%%%%%%%%%%%%%%%%%%%%%%%%%%%%%%%%%%%%%%%%%%%%
%%%%%%%%%%%%%%%%%%%%%%%%%%%%%%%%%%%%%%%%%%%%%%%%%%%%%%%%
\section{The setup for a curved brane model}
Our braneworld model consists of a 3-brane embedded in a five-dimensional
bulk space $\mathcal{M}\times\mathbb{R}$ filled with an analogue of a perfect
fluid with equation of state $P=\gamma \rho$, where the `pressure' $P$ and the `density' $\rho$ are
functions only of the fifth dimension, $Y$. The bulk metric is of the form
\be
\label{warpmetric}
g_{5}=a^{2}(Y)g_{4}+dY^{2},
\ee
where $g_{4}$ is the four-dimensional de Sitter or anti de Sitter metric,
{\it i.e.},
\be
\label{branemetrics}
g_{4}=-dt^{2}+f^{2}_{k}g_{3},
\ee
with
\be
\label{g_3}
g_{3}=dr^{2}+h^{2}_{k}g_{2},
\ee
and
\be
\label{g_2}
g_{2}=d\theta^{2}+\sin^{2}\theta d\varphi^{2},
\ee
where $f_{k}=\cosh (H t)/H$ or $\cos (H t)/H $ ($H^{-1}$ is the de Sitter (or AdS)
curvature radius) and $ h_{k}=\sin r$ or $\sinh r $, respectively.

The metric (\ref{warpmetric}) is a warped product on
$\mathbb{R}\times_{a}\mathcal{M}$ the warping factor being positive $a(Y)>0$, it may be considered
as a generalization of the standard Riemannian cone metric $dS^{2}=dY^{2}+Y^{2}g_{4}$ on
$\mathbb{R}\times_{a}\mathcal{M}$ \cite{peterson}, \cite{o'neill}.
The bulk fluid has an energy-momentum tensor of the form
\be
\label{T old}
T_{AB}=(\rho+P)u_{A}u_{B}-Pg_{AB},
\ee
where $A,B=1,2,3,4,5$ and $u_{A}=(0,0,0,0,1)$, with the 5th coordinate corresponding to $Y$.
The five-dimensional Einstein equations,
\be
G_{AB}=\kappa^{2}_{5}T_{AB},
\ee
can be written in the following form:
\bq
\label{syst2i}
\frac{a''}{a}&=&-2 A\frac{(1+2\gamma)}{3}\rho, \\
\label{syst2iii} \frac{a'^{2}}{a^{2}}&=&\frac{2 A}{3}
\rho+\frac{k H^{2}}{a^{2}},
\eq
where $A=\kappa_{5}^{2}/4$, $k=\pm 1$, and the prime $(\,')$ denotes differentiation with
respect to $Y$. The equation of conservation,
\be
\nabla_{B}T^{AB}=0,
\ee
becomes
\be
\label{syst2ii}
\rho'+4(1+\gamma)\frac{a'}{a}\rho=0.
\ee
Integration of the continuity equation (\ref{syst2ii}) gives the following relation between the density
and the warp factor,
\be
\label{rho to a}
\rho= c_{1} a^{-4(\gamma+1)},
\ee
where $c_{1}$ is an arbitrary integration constant. Substitution of Eq.~(\ref{rho to a}) in
Eq.~(\ref{syst2iii}), gives
\be
a'^2=\dfrac{2}{3}A  c_{1}a^{-2(2\gamma+1)}+kH^2,
\ee
and after setting $C=2/3Ac_{1}$ (note the the sign of $C$ is the same with the sign of $\rho$),
we have the Friedman constraint in the form,
\be
\label{integration eq}
a'^2=Ca^{-2(2\gamma+1)}+kH^2.
\ee
The left hand side of this equation restricts the signs of $C$ and $k$, as well as the range of $a$. As a
result, the case $C<0$ and $k<0$ becomes automatically impossible.

On the other hand, the case $C<0$, $k>0$ is possible only for,
\begin{equation}
\label{bound ds}
\textrm{dS braneworld:}\quad
0<a^{-2(2\gamma+1)}<-\dfrac{kH^2}{C},
\end{equation}
while, the case $C>0$, $k<0$ is possible only for
\begin{equation}
\label{bound ads}
\textrm{AdS braneworld:}\quad a^{-2(2\gamma+1)}>-\dfrac{kH^2}{C}>0.
\end{equation}
It is straightforward to see that these two cases offer the possibility for
avoidance of singularities. More generally, our solutions below are
characterised by these three constants, namely, the curvature constant $k$, the fluid constant
$\gamma$, and the constant $C$, where as mentioned above the sign of $C$ controls that of the density $\rho$.
%We know from previous work \cite{ack2} in which we performed an asymptotic analysis of the dynamical system
%(\ref{syst2i}), (\ref{syst2iii}) and (\ref{syst2ii}), that the cases $C>0$, $k<0$ and $C<0$, $k>0$
%are non-singular for $-1<\gamma<-1/2$ and $\gamma>-1/2$ respectively.

The case $C<0$, $k>0$ of a dS brane with $\gamma>-1/2$, implies that $a^{2(2\gamma+1)}>-C/(kH^{2})>0$, so that the warp factor, $a$, is bounded away from zero, excluding therefore
collapse singularities from happening. The only way that this case may introduce a finite-distance
singularity is to have a warp factor that becomes divergent within a finite distance (big-rip singularity).
However, it will follow from our analysis in the next Section that this behaviour is also excluded
and therefore this case does indeed lead to the avoidance of finite-distance singularities.
On the other hand, the case $C>0$, $k<0$ of an AdS brane with $\gamma<-1/2$, implies that the warp factor
takes only values greater than $-kH^2/C>0$, thus excluding the existence of collapse
singularities, as well.  As we will show later on, this latter case requires a further restriction on
$\gamma$, $-1<\gamma<-1/2$, in order to avoid a finite-distance big-rip singularity.
%%%%%%%%%%%%%%%%%%%%%%%%%%%%%%%%%%%%%%%%%%%%%%%%
%%%%%%%%%%%%%%%%%%%%%%%%%%%%%%%%%%%%%%%%%%%%%%%%
\section{Analysis of the Friedman equation}
Eq. (\ref{integration eq}) can be integrated out to give a solution represented by the Gaussian
hypergeometric function $_{2}F_{1}$, for all possible cases defined by the signs
of $C$ and $k$ and the range of $\gamma$
\footnote{The classification of all the possible cases studied in Sections 3.1-3.3 is defined
through the restriction of obtaining a valid integral representation of the Gaussian
hypergeometric function from Eq.~(\ref{integration eq}), \cite{wang}.}. These solutions along with the asymptotic
behaviours they introduce are presented in what follows. The values $\gamma=-1$ and $\gamma=-1/2$ are special values for our system of equations
(\ref{syst2i}), (\ref{syst2iii}) and (\ref{syst2ii}) in the sense that they reduce it to a significantly
simpler form which leads to solutions that cannot be incorporated into the solutions found below with the use of the
Gaussian hypergeometric function. The solutions for these values of $\gamma$ are studied separately
in Appendix A, however, in order to have a complete view of the asymptotics for
all possible values of $\gamma$ immediately, we present the behaviors of the solutions for $\gamma=1,-1/2$
in the Subsections 3.1, 3.2 and 3.3 that follow.
%%%%%%%%%%%%%%%%%%%%%%%%%%%%%%%%%%%%%%%%%%%%%%%%%%%%%%%%
%%%%%%%%%%%%%%%%%%%%%%%%%%%%%%%%%%%%%%%%%%%%%%%%%%%%%%%%
\subsection{dS branes with positive density}
We begin our study with the case of a dS braneworld with positive density.
This corresponds to $C>0$, $k>0$, and depending on the range of $\gamma$ we have
the following sub-cases:
\begin{itemize}
\item[Ia)] For $\gamma<-1/2$, the solution is,
\be
\label{sol type Ia}
\pm (Y-Y_{0})=\dfrac{a}{\sqrt{kH^2}}\,_{2}F_{1}\left(\dfrac{1}{2},-\dfrac{1}{2(2\gamma+1)},
\dfrac{4\gamma+1}{2(2\gamma+1)},-\dfrac{C}{kH^2}a^{-2(2\gamma+1)}\right).
\ee

\item[Ib)] For $\gamma>-1/2$, we have,
\be
\label{sol type Ib}
\pm (Y-Y_{0})=\dfrac{1}{2(\gamma+1)\sqrt{C}}a^{2(\gamma+1)}\,
_{2}F_{1}\left(\dfrac{1}{2}, \dfrac{\gamma+1}{2\gamma+1},\dfrac{3\gamma+2}{2\gamma+1},
-\dfrac{k H^2}{C}a^{2(2\gamma+1)}\right).
\ee
\end{itemize}
The possible asymptotic behaviours follow from those of the hypergeometric function $_{2}F_{1}$:
\begin{enumerate}
  \item [$\bullet$] $\gamma\geq-1/2$ we find that $a\rightarrow 0$, as
  $Y\rightarrow Y_{0}$. This is a \emph{collapse} type singularity and it
  appears within a finite distance, at $Y_{0}$.
  \item [$\bullet$] $\gamma\geq-1/2$ we have that $a\rightarrow \infty$, as
  $Y\rightarrow\infty$, which describes the behaviour of the warp factor at infinite distance.
  \item [$\bullet$] $\gamma\leq-1/2$ the behaviour here is $a\rightarrow 0$, as
  $Y\rightarrow Y_{0}$, so that we have a collapse singularity at $Y_{0}$.
  \item [$\bullet$] $-1\leq\gamma\leq-1/2$ we get $a\rightarrow\infty$, as
  $Y\rightarrow\infty$, which is as before the behaviour of the warp factor at infinite distance.
  \item [$\bullet$] $\gamma<-1$, in this case $a\rightarrow\infty$, as
  $Y\rightarrow Y_{0}\pm \delta$,
  where the constant $\delta$ is given by:
  \be
  \delta=\dfrac{\Gamma((1+4\gamma)(2(2\gamma+1))\Gamma((\gamma+1)/(2\gamma+1))}
  {\sqrt{\pi} (kH^2)^{\frac{\gamma+1}{2\gamma+1}}}C^{\frac{1}{2(2\gamma+1)}}
  \ee
  where $\Gamma$ is the Gamma function.
This is a \emph{big rip} singularity and it appears within finite distance, at
  $Y_{0}\pm \delta$.
  %\item [$\bullet$] $\gamma=-1$, in this case $a\rightarrow 0$, as
  %$Y\rightarrow \dfrac{\sqrt{6}}{2\kappa_{5}\sqrt{c_{3}}}\ln \left(-\dfrac{c_{2}}{c_{1}}\right)$,
%where $c_{1}$, $c_{2}$ and $c_{3}$ are arbitrary constants (see Appendix A).
\end{enumerate}
For $\gamma<-1$ there are two types of finite-distance singularities: a collapse
singularity located at $Y_{0}$, \emph{and} a big-rip singularity located at
$Y_{0}\pm \delta$. As we will show later, the coexistence of these two types of singularity
not only does not lead to non-singular spacetimes, but it also impedes the
construction of any non-singular matching solution for $\gamma<-1$.

On the other hand, for $\gamma\geq -1$, the braneworld suffers only from a finite-distance
singularity of the collapse type, which allows for the construction of a matching non-singular
solution.
%%%%%%%%%%%%%%%%%%%%%%%%%%%%%%%%%%%%%%%%%%%%%%%%%%%%%
%%%%%%%%%%%%%%%%%%%%%%%%%%%%%%%%%%%%%%%%%%%%%%%%%%%%%
\subsection{dS branes with negative density}
We now consider a dS braneworld with negative density, corresponding to
$C<0$, $k>0$, and let $\gamma$ varying as follows:
\begin{itemize}
\item [IIa)] $\gamma<-1/2$, the solution is given by (\ref{sol type Ia}).

\item [IIb)] $\gamma>-1/2$, the solution is,
\beq
\nonumber
\label{sol type IIb}
\pm (Y-Y_{0})&=&\dfrac{(-C)^{-\frac{\gamma}{2\gamma+1}}(kH^2)^{\frac{-1}{2(2\gamma+1)}}}
{2(2\gamma+1)}\sqrt{a^{2(2\gamma+1)}+\dfrac{C}{kH^2}}\times\\
& &\times _{2}F_{1}\left(\dfrac{1}{2},\dfrac{\gamma}{2\gamma+1},\dfrac{3}{2},
1+\dfrac{k H^2}{C}a^{2(2\gamma+1)}\right).
\eeq
\end{itemize}
The solutions in this case satisfy the bounds (\ref{bound ds})
%\be
%\label{cond a sol II}
%0<a^{-2(2\gamma+1)}<-\dfrac{kH^2}{C}
%\ee
and we arrive at the following asymptotic behaviours:
\begin{enumerate}
  \item [$\bullet$] $\gamma\leq-1/2$ here $a\rightarrow 0$ as $Y\rightarrow Y_{0}$, and
  this is a collapse singularity appearing at $Y_{0}$. There is no other singularity since $a$
is bounded from above and never diverges.

 \item [$\bullet$] $\gamma>-1/2$ we see that $a\rightarrow\infty$ as
  $Y\rightarrow\infty$, which means that this case is free from finite-distance singularities,
  since $a$ is bounded from below and never vanishes.
\end{enumerate}
%%%%%%%%%%%%%%%%%%%%%%%%%%%%%%%%%%%%%%%%%%%%%%%%%
%%%%%%%%%%%%%%%%%%%%%%%%%%%%%%%%%%%%%%%%%%%%%%%%%
\subsection{AdS branes with positive density}
The last possible case is that of an open universe with positive density support on the bulk which
translates to considering $C>0$ and $k<0$ (AdS braneworld). Taking into account the possible ranges
of $\gamma$ we have the following outcomes:
\begin{itemize}
\item[IIIa)] $\gamma>-1/2$ or, $\gamma<-1$ the solution is
given by Eq. (\ref{sol type Ib}).

\item[IIIb)] $-1<\gamma<-1/2$ the solution is,
\beq
\nonumber
\label{sol type IIIb}
\pm (Y-Y_{0})&=&-\dfrac{C^{\frac{\gamma+1}{2\gamma+1}}}{2(2\gamma+1)(-k H^2)^
{\frac{4\gamma+3}{4\gamma+2}}}
\sqrt{a^{-2(2\gamma+1)}+\dfrac{kH^2}{C}}\,\times\\
&\times&  _{2}F_{1}
\left(\dfrac{4\gamma+3}{2(2\gamma+1)},\dfrac{1}{2}, \dfrac{3}{2},
1+\dfrac{C}{k H^2}a^{-2(2\gamma+1)}\right).
\eeq
\end{itemize}
As we mentioned earlier, this case is subject to the bound (\ref{bound ads})
%\be
%\label{cond a sol III}
%a^{-2(2\gamma+1)}>-\dfrac{kH^2}{C}>0.
%\ee
The possible asymptotic behaviours for this case are then as follows:
\begin{enumerate}
  \item [$\bullet$] $\gamma\geq-1/2$ we find $a\rightarrow 0$, as
  $Y\rightarrow Y_{0}$, which implies a collapse singularity at
  $Y_{0}$; the warp factor is bounded from above.
  \item [$\bullet$] $-1\leq\gamma<-1/2$  we find $a\rightarrow\infty$ as
  $Y\rightarrow\infty$, so that this region of $\gamma$ is free form finite distance singularities,
  since $a$ is again bounded from below and never vanishes.
  \item [$\bullet$] $\gamma<-1$ we have $a\rightarrow\infty$, as
  $Y\rightarrow Y_{0}$. This is a big-rip singularity located at $Y_{0}$.
  There is no collapse singularity since $a$ is bounded from below.
\end{enumerate}
%%%%%%%%%%%%%%%%%%%%%%%%%%%%%%%%%%%%%%%%%%%%%%%%%%%%%%%
%%%%%%%%%%%%%%%%%%%%%%%%%%%%%%%%%%%%%%%%%%%%%%%%%%%%%%%
\section{Non-singular solutions}
We saw in the previous Section that there are solutions free from finite-distance singularities in the following cases:
\begin{itemize}
\item dS brane with negative density and $\gamma> -1/2$

\item AdS brane with positive density and $-1\leq\gamma<-1/2$.
\end{itemize}
In this Section we analyse the complete character of these two non-singular solutions.% (Case $II$ and Case $III$, respectively).

The first non-singular solution is given by Eq.~(\ref{sol type IIb}) for $\gamma>-1/2$. The two branches of this solution
may be matched in the following way. Let $\hat{Y}$ denote the value of $Y$ for which
$a^{-2(2\gamma+1)}$ is equal to $-kH^{2}/C$, that is
\be
\label{a hat}
(a(\hat{Y}))^{-2(2\gamma+1)}=-\dfrac{kH^{2}}{C}.
\ee
Then we see from Eq.~(\ref{sol type IIb}) that the value of the Gaussian hypergeometric function
at $\hat{Y}$ is equal to one, while from the left-hand side of the same equation we find that $\hat{Y}=Y_{0}$.
We note that $Y_{0}$ is a regular point of the solution. Putting Eq.~(\ref{a hat}) in
Eq.~(\ref{integration eq}) we find that
\be
a'(Y_{0})=0,
\ee
so that $Y_{0}$ is a critical point. We now check the second derivative of $a$, $a''$.
For $C<0$ it follows from Eq.~(\ref{rho to a}) and the fact that $c_{1}=3C/(2A)$
that also $\rho<0$. Further assuming $\gamma>-1/2$, we find from Eq~(\ref{syst2i}) that
\be
a''(Y_{0})>0.
\ee
Thus, for this choice of parameters we see that at $Y_{0}$ the warp
factor takes its minimum value and then it starts to increase, avoiding in this way collapse
singularities.

\begin{figure}
  \centering
  \includegraphics[scale=0.5]{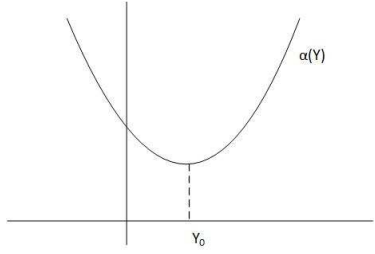}\\
  \caption{\it{\small{Non-singular solution for $C<0$, $k>0$ and $\gamma>-1/2$,
  or $C>0$, $k<0$ and $-1<\gamma<-1/2$.}}}\label{non singular solution}
\end{figure}

To study the behaviour of the solution at infinity i.e. as $a\rightarrow\infty$, we expand the
hypergeometric function as follows \cite{wang},
\beq
\nonumber
& &_{2}F_{1}\left(\dfrac{1}{2},\dfrac{\gamma}{2\gamma+1},\dfrac{3}{2},
1+\dfrac{k H^2}{C}a^{2(2\gamma+1)}\right)=
\Gamma_{1}\left(-1-\dfrac{kH^2}{C}a^{2(2\gamma+1)}\right)^{-1/2}\times
\\
\nonumber
&\times&_{2}F_{1}\left(\frac{1}{2},0,\frac{4\gamma+3}{2(2\gamma+1)},
\left(1+\frac{kH^2}{C}a^{2(2\gamma+1)}\right)^{-1}\right)
+\Gamma_{2}\left(-1-\frac{kH^2}{C}a^{2(2\gamma+1)}\right)^{-\frac{\gamma}{2\gamma+1}}\times
\\
&\times& _{2}F_{1}\left(\frac{\gamma}{2\gamma+1}, -\frac{1}{2(2\gamma+1)},\frac{4\gamma+1}{2(2\gamma+1)},
\left({1+\frac{kH^2}{C}a^{2(2\gamma+1)}}\right)^{-1}\right),
\eeq
where
\be
\Gamma_{1}=\frac{\sqrt{\pi}\,\Gamma(-1/(2(2\gamma+1)))}{2\Gamma(\gamma/(2\gamma+1))}
\ee
and
\be
\Gamma_{2}=\frac{\Gamma(1/(2(2\gamma+1)))}{2\Gamma((4\gamma+3)/(2(2\gamma+1)))}.
\ee
Substituting in the solution (\ref{sol type IIb}) we find,
\be
\label{asympt 1}
\pm(Y-Y_{0})\sim a,
\ee
and so we see from (\ref{asympt 1}) that $a\rightarrow \infty$ is only possible for
$Y\rightarrow\pm\infty$. The behaviour of $a$ is shown in
Fig. \ref{non singular solution}.

For an AdS braneworld on the other hand, the solution (\ref{sol type IIIb}) is non-singular
for $-1<\gamma<-1/2$. We note that because of Eq.~(\ref{bound ads}), the warp factor $a$ for this range of $\gamma$ cannot
approach zero and therefore all collapse singularities are excluded from happening, however,
the possibility of the warp factor becoming \emph{divergent} within finite distance is not a priori
prohibited. If such behaviour is encountered then we end up with an even stronger type of
singularity, a big rip.

Let us suppose that the warp factor does become divergent,
$a\rightarrow \infty$, but restrict $\gamma$ in the interval $(-1,-1/2)$. The hypergeometric
function appearing in Eq.~(\ref{sol type IIIb}) has the argument
$1+C a^{-2(2\gamma+1)}/(kH^{2})$ which is diverging so to study its behaviour at infinity, we
first expand it in the following way,
\beq
\nonumber
& &_{2}F_{1}\left( \dfrac{4\gamma+3}{2(2\gamma+1)},\dfrac{1}{2},\dfrac{3}{2},
1+\dfrac{C}{k H^2}a^{-2(2\gamma+1)}\right)=\\
\nonumber
& &\Gamma_{3}\left(-1-\dfrac{C}{kH^2}a^{-2(2\gamma+1)}\right)^{-\frac{(4\gamma+3)}{4\gamma+2}}
\times\\
\nonumber
& &\times_{2}F_{1}\left(\dfrac{4\gamma+3}{2(2\gamma+1)}, \dfrac{\gamma+1}{2\gamma+1},
\dfrac{3\gamma+2}{2\gamma+1}, \left(1+\dfrac{C}{kH^2}a^{-2(2\gamma+1)}\right)^{-1}\right)+\\
& & +\Gamma_{4}\left(-1-\dfrac{C}{kH^2}a^{-2(2\gamma+1)}\right)^{-1/2}\,
_{2}F_{1}\left(\dfrac{1}{2},0,\dfrac{\gamma}{2\gamma+1},
\left(1+\dfrac{C}{kH^2}a^{-2(2\gamma+1)}\right)^{-1}\right),
\eeq
where $\Gamma_{3}$ and $\Gamma_{4}$ are the constants,
\be
\Gamma_{3}=\dfrac{\Gamma(-(\gamma+1)/(2\gamma+1))}{2\Gamma(\gamma/(2\gamma+1))},
\ee
and
\be
\Gamma_{4}=\dfrac{\sqrt{\pi}\,\Gamma((\gamma+1)/(2\gamma+1))}
{2\Gamma((4\gamma+3)/(2(2\gamma+1)))}.
\ee
Substituting the above expression of the hypergeometric function in the solution
(\ref{sol type IIIb}), we deduce the asymptotic behaviour,
\be\label{asyminfty}
\pm(Y-Y_{0})\sim a^{2(\gamma+1)},
\ee
so that for $a\rightarrow\infty$, we get $Y\rightarrow\pm\infty$.
Therefore the divergence of the warp factor is only possible at infinite distance which means
that finite-distance big rip singularities are excluded. The behaviour of $a$ is similar to the
one shown in Fig. \ref{non singular solution}.
%For $\gamma<-1$ on the other hand, (\ref{asyminfty}) implies that $a$ diverges at $Y=Y_0$, corresponding to a big rip singularity.
%%%%%%%%%%%%%%%%%%%%%%%%%%%%%%%%%%%%%%%%%%%%%%%%%%%%%%%
%%%%%%%%%%%%%%%%%%%%%%%%%%%%%%%%%%%%%%%%%%%%%%%%%%%%%%%
\section{The null energy condition}
In this Section we study the null energy condition for our type of matter (\ref{T old}) and then
examine for which ranges of $\gamma$ it holds true.

We note that our metric (\ref{warpmetric}) and our fluid are static with respect to the time coordinate $t$.
%First, we rewrite (\ref{T old}) in the form
%\be
%T_{AB}^{\textrm{old}}=(\rho_{\textrm{old}}+p_{\textrm{old}})u_{A^{old}}u_{B}^{\textrm{old}}-
%-p_{\textrm{old}}g__{AB},
 %\ee
%where $u_{A^{old}}=(0,0,0,0,1)$ and transform $T_{AB}^{\textrm{old}}$ into a new energy
%momentum tensor, which we denote by $T_{AB}^{\textrm{new}}$, that follows once may
%reinterpret our fluid as an anisotropic fluid:
We may reinterpret our fluid analogue as a real anisotropic fluid having the following energy
momentum tensor:
\be
\label{T new}
T_{AB}= (\rho^{0}+
p^{0})u_{A}^{0}u_{B}^{0}
+p^{0}g_{\alpha\beta}\delta_{A}^{\alpha}\delta_{B}^{\beta}+
%p_{Y}g_{YY}\delta_{A}^{Y}\delta_{B}^{Y},
p_{Y}g_{55}\delta_{A}^{5}\delta_{B}^{5},
\ee
where $u_{A}^{0}=(a(Y),0,0,0,0)$, $A,B=1,2,3,4,5$ and $\alpha,\beta=1,2,3,4$.
When we combine (\ref{T old}) with (\ref{T new}), we get the following set of relations,
\bq
\label{p y to rho}
p_{Y}&=&\rho\\
\label{rho new}
\rho^{0}&=&p\\
\label{p new}
p^{0}&=&-p.
\eq
The last two relations imply that
\be
\label{p new to rho new}
p^{0}=-\rho^{0},
\ee
which means that this type of matter satisfies a cosmological constant-like equation of state.
Imposing further $p=\gamma\rho$, and using (\ref{p y to rho}), leads to,
\be
\label{py to p}
p_{Y}=\frac{p}{\gamma}.
\ee
Substituting (\ref{p new}), (\ref{p new to rho new}) and (\ref{py to p}) in (\ref{T new}), we
find that
\be
T_{AB}= -p g_{\alpha\beta}\delta_{A}^{\alpha}\delta_{B}^{\beta}+
%\frac{p}{\gamma}g_{YY}\delta_{A}^{Y}\delta_{B}^{Y}.
\frac{p}{\gamma}g_{55}\delta_{A}^{5}\delta_{B}^{5}.
\ee
We are now ready to form the null energy condition for our type of matter.
According to the null energy condition, every future-directed
null vector $k^{A}$ should satisfy \cite{poisson}
\be
T_{AB}k^{A}k^{B}\geq 0.
\ee
This condition implies that the energy density should be non-negative. Here we find that it translates to
\be
p+\dfrac{p}{\gamma}\geq 0,
\ee
or, in terms of $\rho$
\be
(\gamma+1)\rho\geq 0,
\ee
which leads to two possible cases, namely,
\be
\rho\geq 0 \quad \textrm{and} \quad \gamma\geq-1,\quad\textrm{or,}\quad
\rho\leq 0 \quad \textrm{and}\quad \gamma\leq-1.
\ee
%\be
%\label{nec_p_1}
%p\geq 0 \quad \textrm{and}\quad  \gamma>0, \quad \textrm{or},\quad  \gamma\leq-1,
%\ee
%or,
%\be
%\label{nec_p_2}
%p\leq 0 \quad \textrm{and} \quad -1\leq\gamma<0.
%\ee
With the use of (\ref{rho to a}), in which $c_{1}=3/(2A)C$, these two conditions may be written equivalently with
respect to $C$ instead of $\rho$ as,
\be
\label{nec_c_1}
C\geq 0 \quad \textrm{and}\quad  \gamma\geq-1,
\ee
and
\be
\label{nec_c_2}
C\leq 0 \quad \textrm{and} \quad \gamma\leq-1.
\ee

The conditions (\ref{nec_c_1}) and (\ref{nec_c_2}) show that the requirement of satisfying the
null energy condition leads to restrictions on both the range of $\gamma$ and the sign of
the constant $C$. We conclude that the only range for $C$ and $\gamma$ that is compatible
with a non-singular solution, and at the same time also satisfies the null energy condition, is
$C>0$ and $-1<\gamma<-1/2$ combined with $k<0$, that is AdS braneworld with positive
density and $\gamma\in(-1,-1/2)$. In particular, dS non-singular braneworlds are incompatible
with the null energy condition holding in the bulk.
%As we will show in the next
%Section the range of $\gamma$ that gives also a finite Planck mass is $(-2,-1)$
%(see (\ref{finite_planck}), which implies that it is possible to satisfy at the same time
%the null energy condition \emph{and} the requirement for a finite Planck mass.
%%%%%%%%%%%%%%%%%%%%%%%%%%%%%%%%%%%%%%%%%%%%%%%%%%%%%%
%%%%%%%%%%%%%%%%%%%%%%%%%%%%%%%%%%%%%%%%%%%%%%%%%%%%%%%
\section{Matching solutions}
In this Section we will examine those solutions from the Sections 3.1-3.3,
that allow a jump in the derivative of the warp factor, $a'$, across the brane, and also
satisfy the null energy condition. These are the cases $Ia)$ and $Ib)$.

For the case $Ia)$, we have $\gamma<-1/2$, and for the null energy condition we should
further restrict to $-1<\gamma<-1/2$. Setting $c_{2}=\mp Y_{0}$, and choosing the $+$ sign of $Y$ for $Y>0$ and the $-$ sign for $Y<0$,
the solution (\ref{sol type Ia}) can be written in the form
\be
\label{matching sol 1}
|Y|+c_{2}^{\pm}=\dfrac{a}{\sqrt{kH^2}}\,_{2}F_{1}\left(\dfrac{1}{2},-\dfrac{1}{2(2\gamma+1)},
\dfrac{4\gamma+1}{2(2\gamma+1)},-\dfrac{C}{kH^2}a^{-2(2\gamma+1)}\right).
\ee
For $c_{2}^{\pm}>0$, we see that collapse singularities are excluded. Since we have restricted
$\gamma$ to take values greater than $-1$, big-rip singularities are also excluded.
Assuming a continuous warp factor at the position of the brane $Y=0$, we get from
(\ref{matching sol 1}) a condition for $c_{2}$ which reads,
\be
\label{match_1}
c_{2}^{+}=c_{2}^{-},
\ee
where $c_{2}^{\pm}$ denote the values of $c_{2}$ at $Y=0^{\pm}$. Note that $c_{2}^{\pm}$ are
both positive from (\ref{matching sol 1}) which is compatible with
our choice of sign for $c_{2}$. Further imposing continuity of $\rho$ at $Y=0$, we find
from (\ref{rho to a}) and $c_{1}=3/(2A)C$ that
\be
\label{match_2}
C^{+}=C^{-}.
\ee
Next, we take into account the jump of the derivative of the warp factor across the brane.
For our type of geometry this junction condition reads
\be
\label{match_3}
a'(0^{+})-a'(0^{-})=-\dfrac{1}{3}f(\rho(0))a(0),
\ee
where $f(\rho)$ is the tension of the brane. For our solution the above condition translates to
\be
\label{match_4}
f(\rho(0))=-6\sqrt{kH^{2}a^{-2}(0)+Ca^{-4(\gamma+1)}(0)},
\ee
from which we note that the brane tension is negative.
\begin{figure}
\label{matching two solution}
  \centering
  \includegraphics[scale=0.5]{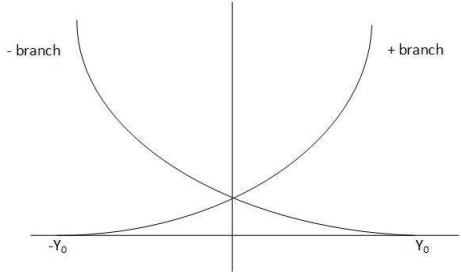}\\
  \caption{\it{\small{Non-singular matching solution for $C>0$, $k>0$ and $-1<\gamma<-1/2$. We take the $(+)$ branch for
  $Y>0$ and the $(-)$ branch for $Y<0$.}}}
\end{figure}
In Fig. $2$, %\ref{matching one solution}
we depict the two branches of the solution and the way they may be matched together to give
the non-singular solution described above.
%
%The case $IIa)$ has the same form of original solution (\ref{sol type Ia}) as the case $Ia)$ that
%we just studied and it can therefore admit the same form of matching solution given by
%Eq.~(\ref{matching sol 1}) and satisfy the same junction conditions
%given by equations (\ref{match_1})-(\ref{match_3}). Note that
%$\gamma<-1/2$ in the case $IIa)$ and in order to satisfy the null
%energy condition we take $\gamma<-1$. Then we can have a matching solution that is
%non-singular shown in Fig. $2$ that also satisfies the null energy condition.

Similarly, we may match the two branches of solution of case $Ib)$ for $\gamma>-1/2$, and
find,
\be
\label{matching sol 2}
|Y|+c_{2}^{\pm}=\dfrac{1}{2(\gamma+1)\sqrt{C}}a^{2(\gamma+1)}\,
_{2}F_{1}\left(\dfrac{1}{2}, \dfrac{\gamma+1}{2\gamma+1},\dfrac{3\gamma+2}{2\gamma+1},
-\dfrac{k H^2}{C}a^{2(2\gamma+1)}\right).
\ee
It follows that conditions (\ref{match_1})-(\ref{match_4})
are also true in this case. The two branches of the solution and the way they may be matched
together to give the non-singular solution are shown in Fig. $3$.
%\ref{matching two solution}.

\begin{figure}
\label{matching one solution}
  \centering
  \includegraphics[scale=0.5]{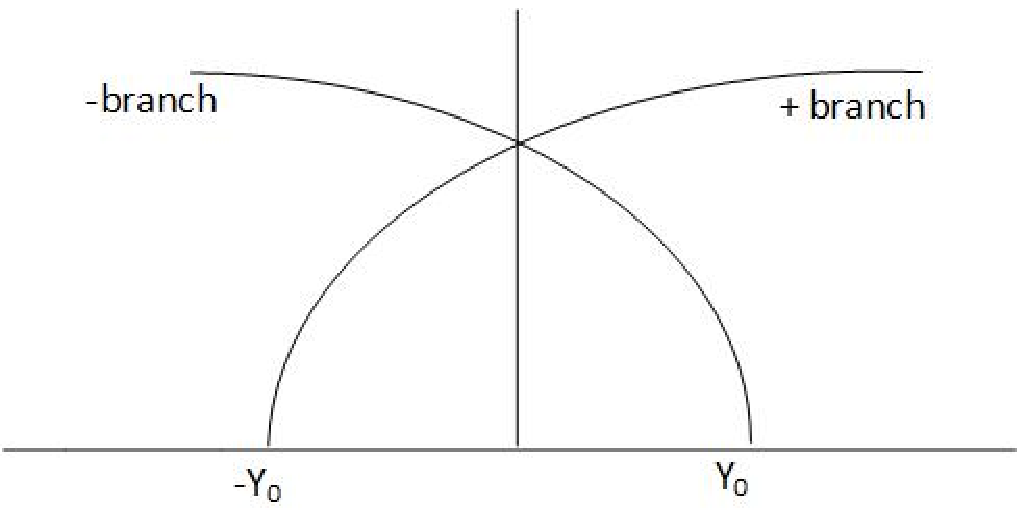}\\
  \caption{\it{\small{Non-singular matching solution for $C>0$, $k>0$ and $\gamma>-1/2$.
We take the $(+)$ branch for $Y>0$ and the $(-)$ branch for $Y<0$.}}}
\end{figure}
%
%The last case $IIIa)$ describes an AdS brane with positive density and it is defined through
%the same original solution (\ref{sol type Ib}) as the case $Ib)$ with either
%$\gamma>-1/2$, or, $\gamma<-1$. Therefore it can take the form
%of the matching solution (\ref{matching sol 2}) and satisfy the junction
%conditions (\ref{match_1})-(\ref{match_3}). If we take $\gamma>-1/2$
%then this solution satisfies also the null energy condition. The resulting
%matching solution is shown in Fig. $2$.
%\begin{figure}
%\label{matching four solution}
 % \centering
  %% Requires \usepackage{graphicx}
  %\includegraphics[scale=0.5]{matching4.jpg}\\
  %%\includegraphics[scale=1.5]{nonsing.pdf}\\
%  \caption{\it{\small{Non-singular matching solution for $C>0$, $k<0$ and $\gamma<-1$.
%We take the $(-)$ branch for $Y>0$ and the $(+)$ branch for $Y<0$.}}}
%\end{figure}
%
%In the next Section, we study the problem of localising gravity on the brane. We show that
%the importance of the solution $IIa)$ is reinforced through this physical demand since
%it is the only solution that can realise the localisation of gravity on the brane.

We can also have a matching solution by cutting the regular solutions $IIb)$, or, $IIIb)$
at a point different from the minimum (cutting the regular solutions at the minimum would
lead to a vanishing brane tension). For example by taking $Y_{0}\neq 0$ and putting the brane at $Y=0$
we can derive the corresponding junction conditions which are again given by
(\ref{match_1})-(\ref{match_3}). The brane tension, however, now reads
\be
f(\rho(0))=-12\sqrt{kH^{2}a^{-2}(0)+Ca^{-4(\gamma+1)}(0)},
\ee
and we note that it is again negative.

The rest of the cases $II$ and $III$ that satisfy the null energy condition for $\gamma<-1$
($IIa$) and $\gamma>-1$ ($IIIa$), respectively, are not suitable for constructing non-singular
matching solutions since they exhibit either two collapse (case $IIIa$), or
two big rip singularities (case $IIIa)$) that restrict $Y$ to take values only between
the interval with endpoints the two finite singularities. That means that the resulting matching
solutions cannot
be extended to the whole real line of $Y$.
%, since for these two cases the derivative of the warp factor is continuous
%everywhere.
%%%%%%%%%%%%%%%%%%%%%%%%%%%%%%%%%%%%%%%%%%%%%%%%%%%%%%
%%%%%%%%%%%%%%%%%%%%%%%%%%%%%%%%%%%%%%%%%%%%%%%%%%%%%%
\section{Localisation of gravity}
Another question we would like to answer is whether the non-singular solutions we have found
for the warp factor $a$, satisfying the null energy condition, lead to a finite four-dimensional
Planck mass, thus localising 4d gravity on the brane for some range of the parameter $\gamma$.
The value of the four-dimensional Planck mass, $M_{p}^{2}=8\pi/\kappa$, is determined by
the following integral \cite{forste},
\be
\frac{\kappa_{5}^{2}}{\kappa}=\int_{-Y_{c}}^{Y_{c}}a^{2}(Y)dY.
\ee
For our first matching solution, Eq.~(\ref{matching sol 1}), the behaviour of $a^{2}$ at large $|Y|$ is
\be
\label{asq linear}
a^{2}\sim (|Y|+c_{2})^{2},
\ee
and the above integral becomes,
\be
\int_{-Y_{c}}^{Y_{c}}(|Y|+c_{2})^{2}dY
=\dfrac{1}{3}\left(-(-Y+c_{2})^{3}|_{-Y_{c}}^{0}+(Y+c_{2})^{3}|_{0}^{Y_{c}}\right).
%&=&\mp\dfrac{1}{2(\gamma+2)}\sqrt{\dfrac{3}{2A c_{1}}}\left(2(\gamma+1)\left(-\sqrt{\dfrac{2Ac_{1}}{3}}|Y|
%+c_{2}\right)\right)^{(\gamma+2)/(\gamma+1)}|_{-Y_{c}}^{Y_{c}},
\ee
%where the $-$ sign occurs for $Y>0$ and the $+$ for $Y<0$.
In the limit $Y_{c}\rightarrow\infty$, the Planck mass becomes infinite.

The same behavior is valid also for the regular solution $IIb)$ found for a dS brane with negative
density and $\gamma>-1/2$. Placing the brane at $Y=0$ and using the line of thinking
of Section $6$ we can bring the left hand side of solution (\ref{sol type IIb}) into a form
involving the absolute value of $Y$. Then Eq.~(\ref{asympt 1}) can take the form of
(\ref{asq linear}) which lead to an infinite Planck mass.

For our second matching solution (\ref{matching sol 2}), the behaviour of $a^{2}$ is
\be
\label{asq gamma}
a^{2}\sim(|Y|+c_{2})^{\frac{1}{\gamma+1}}.
\ee
Integration of $a^{2}$ gives an expression with $Y$ raised to the exponent,
\be
\dfrac{\gamma+2}{\gamma+1}
\ee
which is positive for this case since here $\gamma>-1/2$.
Therefore we see that the Planck mass is infinite also in this case.

As before the same behavior is valid for the other regular solution $IIIb)$ of an AdS brane with positive
density and $-1<\gamma<-1/2$. For this solution we could also place the brane at $Y=0$ and by using
its asymptotic behavior given by Eq.~(\ref{asyminfty}) we see that it could take the form of
(\ref{asq gamma}) and therefore lead to an infinite Planck mass since $\gamma$ does not lie
in $(-2,-1)$.
%For our third matching solution for a dS negative density braneworld and for $\gamma<-1$
%we see that the localisation of gravity is possible. Note that this solution satisfied also the
%null energy condition. This is the unique solution found that satisfies at the same time both
%physical demands.
%
%The matching solution found for AdS brane with positive density has a $\gamma>-1/2$
%and therefore fails to localise gravity on the brane.
%
%Finally there is one more case of a matching solution which we did not study so far because it did
%not satisfy the null energy condition that is the case of an AdS brane with positive density and $\gamma<-1$.
%The matching slution could
%only for
%\be
%\label{finite_planck}
%-2<\gamma<-1,
%\ee
%and takes the form
%\be
%\frac{\kappa_{5}^{2}}{\kappa}=\sqrt{\dfrac{3}{2A c_{1}}}
%\dfrac{(2(\gamma+1)c_{2})^{\frac{\gamma+2}{\gamma+1}}}{
%\gamma+2}.
%+\dfrac{1}{\sqrt{c_{1}^{-}}}\right).
%\ee
 %%%%%%%%%%%%%%%%%%%%%%%%%%%%%%%%%%%%%%%%%%%%%%%%%%%%%%
%%%%%%%%%%%%%%%%%%%%%%%%%%%%%%%%%%%%%%%%%%%%%%%%%%%%%%
\section{Conclusions}
In this paper, we have analysed braneworld singularities in the presence of dS or AdS
branes and found one non-singular solution for a dS brane with negative (bulk) density
and another one for an AdS brane with positive (bulk) density, for particular ranges of the
parameter space, that we constructed explicitly. As we showed in \cite{ack3} this was
impossible for Minkowski branes. In the case of AdS branes the null energy condition is also satisfied.

Comparing and contrasting the results of the asymptotic behaviour of the solutions found in this
paper to those of our previous work \cite{ack2} which was implemented with a different
method of asymptotic analysis we extract the following conclusions: The case of a dS brane with positive
density, described in Section 3.1, was asymptotically constructed by the two balances simultaneously
$_{\gamma}\mathcal{B}_{1}$ and $_{\gamma}\mathcal{B}_{2}$. The first balance
described the behavior of $a$ around a finite collapse singularity, while the second balance
described the behavior of $a$ at infinity. In particular, the balance
$_{\gamma}\mathcal{B}_{2}$ gave the behavior, $a\rightarrow\infty$ as $Y\rightarrow\infty$, however,
this case is characterised as singular because of the finite-distance
singularity of collapse type introduced by the balance $_{\gamma}\mathcal{B}_{1}$.

On the other hand, the case of an AdS brane with positive density, described in Section 3.2,
was depicted by the balance $_{\gamma}\mathcal{B}_{2}$
which is non-singular for $\gamma>-1/2$. The balance
$_{\gamma}\mathcal{B}_{1}$ which leads to finite-distance singularities for flat and positively
curved branes is not valid in this case since it assumes only positive
density, whereas, this case is characterised by a negative density.
Lastly, the third case, described in Section 3.3, can be described by the balance
$_{\gamma}\mathcal{B}_{1}$ which allows for a non-singular solution for $-1<\gamma<-1/2$.

It is possible that non-singular solutions that satisfy the null energy condition in
the bulk and at the same time localize gravity in the braneworld exist for models of interacting matter
as in \cite{ack4}, or, for homogeneous but anisotropic
(eg. Bianchi I, V, or VIII, IX) braneworlds. Exploring such models will help us decide about the stability of our
non-singular solutions discovered here with respect to more general (anisotropic) perturbation.
We leave this to a future publication.
%%%%%%%%%%%%%%%%%%%%%%%%%%%%%%%%%%%%%%%%%%%%%%%%%%%%%%
%%%%%%%%%%%%%%%%%%%%%%%%%%%%%%%%%%%%%%%%%%%%%%%%%%%%%%
\section*{Acknowledgements}
The authors would like to thank an anonymous referee of \cite{ack3} for suggesting and giving a first analysis
on the problem of curved braneworlds studied in this paper. I.K. is grateful to LPTHE for making her
visit there possible and for financial support.
%%%%%%%%%%%%%%%%%%%%%%%%%%%%%%%%%%%%%%%%%%%%%%%%%%%%%%
%%%%%%%%%%%%%%%%%%%%%%%%%%%%%%%%%%%%%%%%%%%%%%%%%%%%%%
\appendix
\section{Appendix: Solutions for curved branes and special fluids}
In this Appendix we analyze the behavior of the system of equations (\ref{syst2i})-(\ref{syst2iii})
and (\ref{syst2ii}) for special values of the parameter $\gamma$ of the fluid. These are values
of $\gamma$ that simplify the dynamical system significantly and lead to solutions that cannot be
incorporated to the solutions found in Section  $3$.
These special values are $\gamma=-1/2$ and $\gamma=-1$.

Consider first $\gamma=-1/2$. Eqs.~(\ref{syst2i})-(\ref{syst2iii})
and (\ref{syst2ii}) for $\gamma=-1/2$ become
\beq
\label{syst g=-1/2 i}
a''&=&0\\
\label{syst g=-1/2 ii}
\dfrac{{a'}^{2}}{a^{2}}&=&\dfrac{\kappa_{5}^{2}}{6}\rho+\dfrac{kH^{2}}{a^{2}}\\
\label{syst g=-1/2 iii}
\rho'+2\dfrac{a'}{a}\rho&=&0.
\eeq
Eq.~(\ref{syst g=-1/2 i}) gives directly
\be
\label{a for g=-1/2}
a=c_{1}Y+c_{2},
\ee
where $c_{1}$ and $c_{2}$ are arbitrary constants. Inputting (\ref{a for g=-1/2}) in
Eq.~(\ref{syst g=-1/2 iii}) we find
\be
\label{rho for g=-1/2}
\rho=\dfrac{{c}_{3}}{(c_{1}Y+c_{2})^{2}},
\ee
where $c_{3}$ is an arbitrary constant. We substitute Eqs.~(\ref{a for g=-1/2}) and
(\ref{rho for g=-1/2}) in Eq.~(\ref{syst g=-1/2 ii}) to derive the relation between the
three arbitrary constants which reads
\be
c_{3}=\dfrac{6}{\kappa_{5}^{2}}({c_{1}}^{2}-kH^{2}).
\ee
The linear solution (\ref{a for g=-1/2}) shows that for $\gamma=-1/2$, $a\rightarrow 0$
at a finite-distance
\be
Y_{0}=-\dfrac{c_{2}}{c_{1}},
\ee
and also $a\rightarrow\infty$ as $Y\rightarrow \infty$. This case therefore suffers from a
finite-distance collapse singularity.

For $\gamma=-1$, on the other hand, the dynamical system  given by
Eqs.~(\ref{syst2i})-(\ref{syst2iii}) and (\ref{syst2ii}) takes the form
\beq
\label{syst g=-1 i}
\dfrac{a''}{a}&=&\dfrac{\kappa_{5}^{2}}{6}\rho\\
\label{syst g=-1 ii}
\dfrac{{a'}^{2}}{a^{2}}&=&\dfrac{\kappa_{5}^{2}}{6}\rho+\dfrac{kH^{2}}{a^{2}}\\
\label{syst g=-1 iii}
\rho'&=&0.
\eeq
Eq.~(\ref{syst g=-1 iii}) implies that
\be
\label{rho g=-1 c_3<0}
\rho=c_{3},
\ee
with $c_{3}$ an arbitrary constant. By substitution of this in Eq.~(\ref{syst g=-1 i})
we get the following second order differential equation with constant coefficients
\be
a''-\kappa_{5}^{2}\dfrac{c_{3}}{6}a=0
\ee
which has the characteristic equation
\be
\lambda^{2}-\kappa_{5}^{2}\dfrac{c_{3}}{6}=0.
\ee
For $c_{3}>0$ the above equation has two distinct real roots
\be
\label{char-eq}
\lambda=\pm \kappa_{5}\sqrt{\dfrac{c_{3}}{6}}
\ee
and so the general solution has the form
\be
\label{gen sol g=-1}
a=c_{1}e^{{\kappa_{5}}\sqrt{c_{3}/6}Y}+c_{2}e^{{-\kappa_{5}}\sqrt{c_{3}/6}Y}, \quad c_{3}>0,
\ee
where $c_{1}$ and $c_{2}$ are arbitrary constants.
Substituting (\ref{gen sol g=-1}) in Eq.~(\ref{syst g=-1 ii}) we find the relation connecting the
three arbitrary constants which reads
\be
c_{3}=-\dfrac{3 kH^{2}}{2c_{1}c_{2}\kappa_{5}^{2}}.
\ee
Since we have taken $c_{3}>0$ we need to have the following restrictions
on the signs of $c_{1}$, $c_{2}$ and $k$
\be
\textrm{either} \quad c_{1}c_{2}<0 \quad \textrm{and} \quad k>0, \quad \textrm{or,}\quad c_{1}c_{2}>0 \quad \textrm{and} \quad k<0.
\ee
For $c_{1}c_{2}<0$ there is a finite-distance singularity at
\be
Y_{0}=\dfrac{\sqrt{6}}{2\kappa_{5}\sqrt{c_{3}}}\ln \left(-\dfrac{c_{2}}{c_{1}}\right).
\ee
We also see from the solution (\ref{gen sol g=-1}) that $a$ becomes infinite only at infinite $Y$.
For $c_{1}c_{2}>0$ and $k<0$, however, we see that the solution is free from finite-distance
singularities.

For $c_{3}<0$, on the other hand, we have a different behavior. Taking
$c_{3}<0$ translates first from Eq.~(\ref{rho g=-1 c_3<0}) to having negative density and then
from Eq. (\ref{syst g=-1 ii}) to allowing only for a dS brane. For $c_{3}<0$ the
characteristic equation has imaginary roots and
the general solution (\ref{gen sol g=-1}) becomes complex. However, we can still obtain a real
general solution from the complex one by imposing real initial conditions. The real general
solution obtained in this way is given by
\be
a=c_{1}\cos \left( \kappa_{5}\sqrt{\dfrac{C_{3}}{6}}Y\right)+
c_{2}\sin \left( \kappa_{5}\sqrt{\dfrac{C_{3}}{6}}Y\right),
\ee
where $C_{3}=-c_{3}>0$. This solution has an infinite number of finite-distance singularities.

%%%%%%%%%%%%%%%%%%%%%%%%%%%%%%%%%%%%%%%%%%%%%%%%%%%%%%
%%%%%%%%%%%%%%%%%%%%%%%%%%%%%%%%%%%%%%%%%%%%%%%%%%%%%%


\begin{thebibliography}{}
\bibitem{ack3}
I. Antoniadis, S. Cotsakis, I. Klaoudatou, \emph{Enveloping branes and braneworld singularities},
Eur. Phys. J. C74 (2014) 3192, [arXiv:hep-th/1406.0611v2].

\bibitem{ack1}
I. Antoniadis, S. Cotsakis, I. Klaoudatou,
\emph{Braneworld cosmological singularities},
Proceedings of MG11 meeting on General Relativity, vol. 3, pp. 2054-2056,
[arXiv:gr-qc/0701033].

\bibitem{ack2}
I. Antoniadis, S. Cotsakis, I. Klaoudatou, \emph{Brane singularities
and their avoidance}, Class. Quant. Grav. 27 (2010) 235018 [arXiv:gr-qc/1010.6175].

\bibitem{gubser}
S.~S.~Gubser, \emph{Curvature singularities: The good, the bad, and the naked},
Adv. Theor. Math. Phys. 4 (2000) 679, [arXiv:hep-th/0002160].

\bibitem{peterson}
P. Peterson, \emph{Riemannian geometry}, Springer 2006.

\bibitem{o'neill}
B. O' Neill, \emph{Semi-Riemannian geometry with applications to relativity}, Academic Press 1983.

\bibitem{wang}
Z. X. Wang, D. R. Guo, \emph{Special Functions}, World Scientific, 1989.

%\bibitem{wald}R. M. Wald, General Relativity, University of Chicago Press, 1984.
%
\bibitem{poisson} E. Poisson, A Relativist's Toolkit, Cambridge University Press, 2004.

\bibitem{forste}S.~Forste, H.~P.~Nilles and I.~Zavala,
\emph{Nontrival Cosmological Constant in Brane Worlds with Unorthodox
Lagrangians}, JCAP {\bf 1107} (2011) 007, [arXiv:hep-th/1104.2570].

\bibitem{ack4}
Antoniadis, I., Cotsakis, S. and Klaoudatou, I., \emph{Brane singularities
with mixtures in the bulk}, Fortschr. Phys. 61 (2013) 20-49.
\end{thebibliography}
\end{document}